\documentclass[prl,aps,twocolumn,showpacs]{revtex4}
\usepackage{graphicx} % include EPS-figures

\begin{document}
\title{Feasibility of a single-parameter description of equilibrium viscous liquid dynamics}
\author{Ulf R. Pedersen, Tage Christensen, Thomas B. Schr{\o}der, and Jeppe C. Dyre}
\affiliation{DNRF centre  ``Glass and Time,'' IMFUFA, Department of Sciences, Roskilde University, Postbox 260, DK-4000 Roskilde, Denmark}
\date{\today}

\begin{abstract}
Molecular dynamics results for the dynamic Prigogine-Defay ratio are presented for two glass-forming liquids, thus evaluating the experimentally relevant quantity for testing whether metastable-equilibrium liquid dynamics to a good approximation are described by a single parameter. For the Kob-Andersen binary Lennard-Jones mixture as well as for an asymmetric dumbbell model liquid a single-parameter description works quite well. This is confirmed by time-domain results where it is found that energy and pressure fluctuations are strongly correlated on the alpha-time scale in the NVT ensemble; in the NpT ensemble energy and volume fluctuations similarly correlate strongly.
\end{abstract}

\pacs{64.70.Pf; 61.20.Lc}

\maketitle

The physics of viscous liquids approaching the glass transition continue to attract attention \cite{GT_ref}. Basic properties like the origins of non-exponential relaxations or of the non-Arrhenius viscosities are still controversial. A question that is not currently actively debated is whether a single  ``order'' parameter is enough to describe glass-forming liquids and the glass transition \cite{note1}. For more than 30 years the consensus has been that with few exceptions more than one parameter is required, a conclusion that scarcely appears surprising given the complexity of glass-forming liquids. 

The prevailing paradigm in glass science regarding ``order'' parameters may be summarized as follows \cite{GT_ref,note1,pri54,dav,gol64,ang76,nem95,don01}. If $\Delta c_p$ is the drop in isobaric specific heat per volume going from liquid to glass, $\Delta\kappa_T$ and $\Delta\alpha_p$ the same changes in isothermal compressibility and isobaric thermal expansion coefficient respectively, and $T_g$ the glass transition temperature, the Prigogine-Defay ratio $\Pi$ is defined by

\begin{equation}\label{1}
\Pi\,=\,\frac{\Delta c_p\Delta\kappa_T}{T_g\left(\Delta\alpha_p\right)^2}\,.
\end{equation}
Davies and Jones in 1952 proved that if there is just one ``order'' parameter, then $\Pi=1$ \cite{dav}. In their formulation, if a liquid is described by a single parameter, its linear relaxations are all simple exponentials \cite{dav}. The vast majority of reported Prigogine-Defay ratios obey $\Pi>1$  \cite{moy76}, which following Davies and Jones is consistent with the observation that linear relaxations are virtually never exponential. The case for requiring more than one parameter got further support from the classical cross-over experiment of Kovacs \cite{kov63} as well as from several other experiments, all showing that glass structure cannot be completely characterized by a single parameter \cite{roe77,sch86}.

Part of this conventional wisdom may nevertheless be challenged. First, note that $\Pi$ is not strictly well defined. This is because to evaluate $\Delta c_p$, etc., one must extrapolate measurements in the glass phase into the liquid region, but the glass phase is not unique and it relaxes with time. Secondly, as shown by Goldstein and by Moynihan and Lesikar \cite{gol64,moy81} (but perhaps not generally appreciated), it is possible to have systems with $\Pi=1$ described by a single parameter with non-exponential dynamics. Finally note that, although it has been conclusively demonstrated that not all aspects of glass structure can be described by a single number, this does not rule out the possibility that a single parameter is sufficient for describing a more limited range of phenomena, e.g., the linear thermoviscoelastic properties of a glass-forming liquid \cite{nie97,spe99,ell07}. It is this possibility we inquire into below, where all simulations were performed in metastable equilibrium with no reference to the glassy phase.

Experiments carried out the last few years by Richert and Weinstein \cite{ric06}, by Ngai, Casalini, Capaccioli, Paluch, and Roland \cite{nga05}, and by our group \cite{dyr03}, do indicate that a single-parameter description may be appropriate for some situations. The direct motivation of the present work is a paper from 2004 by Mossa and Sciortino who simulated aging of a molecular model of ortho-terphenyl \cite{mos04}. For small temperature steps they found that the location of the aging system in configuration space can be traced back to equilibrium states, implying that \cite{mos04} ``a thermodynamic description based on one additional parameter can be provided'' for cases of  non-linear relaxations fairly close to equilibrium. This suggests studying the metastable equilibrium viscous liquid itself in order to investigate whether -- and how -- ``one-parameter-ness'' may be reflected in the {\it equilibrium fluctuations}.

A single-parameter description of the metastable-equilibrium liquid dynamics by definition applies in the following situation \cite{ell07}. Suppose a linear-response experiment is performed where ``input'' infinitesimal temperature and pressure variations $\delta T(t)$ and $\delta p(t)$ are imposed on the liquid and ``output'' infinitesimal volume and entropy variations $\delta V(t)$ and  $\delta S(t)$ are observed. Since highly viscous liquids exhibit time-scale separation between the fast (vibrational) degrees of freedom and the much slower (relaxational) degrees of freedom, one expects that each of the outputs may be written as instantaneous couplings to the inputs plus a relaxing contribution -- the latter being a standard linear convolution integral involving the previoius temperature and pressure histories. By definition, if the two relaxing contributions are proportional, the liquid is described by a single parameter \cite{ell07}. In other words, a single-parameter description applies whenever a variable $\delta\varepsilon$ exists such that

\begin{eqnarray}\label{order}
 \delta  S(t) &=&  \delta\varepsilon(t) +J_{11}^\infty \delta T(t)   -
 J_{12}^\infty  \delta p(t) \nonumber\\ 
\delta V(t)  &=&  \gamma\, \delta\varepsilon(t) + J_{21}^\infty \delta T(t) 
 -J_{22}^\infty \delta p(t)  \,. 
\end{eqnarray} 

In Ref. \cite{ell07} it was proved that in any stochastic dynamics a single-parameter description applies if and only if the following ``dynamic Prigogine-Defay ratio'' is unity:

\begin{equation}\label{pdf}
\Lambda_{Tp}(\omega)\,=\,
\frac{c''_p(\omega)\kappa''_T(\omega)}{T\left(\alpha''_p(\omega)\right)^2}\,,
\end{equation}
where the double primes denote the imaginary part of the response function. Moreover, it was proved that if the dynamic Prigogine-Defay ratio is unity at one frequency, it is unity at all frequencies. The relevant frequency-dependent thermoviscoelastic response functions are very difficult to measure -- in fact no reliable measurements appear yet to exist. This single-frequency criterion could become useful for the interpretation of experiment, since it is likely that the first measurements will cover only a limited frequency range. In real experiments one can of course never prove that a number is exactly unity, but it seems reasonable to assume that the closer the dynamic Prigogine-Defay ratio is to unity, the more accurate a single-parameter description is. 

There is no simple physical interpretation of the parameter $\delta\varepsilon$ except that it controls both entropy and volume relaxations. In particular, there is no reason to believe that  $\delta\varepsilon$ gives a complete characterization of the molecular structure in the way that the ``order'' parameters of traditional glass science do. Despite this severe limitation it is not obvious that glass-forming liquids exist that are described by a single parameter to a good approximation. Recent results studying thermal fluctuations of less-viscous liquids give rise to optimism, though: For a number of model liquids \cite{ped07} -- including the standard Lennard-Jones system -- the potential energy and the virial correlate better than 90\% in their thermal equilibrium fluctuations. Recall that the virial (when divided by volume) is the contribution to pressure from the molecular interactions, i.e., in addition to the ever-present ideal-gas contribution deriving from the kinetic degrees of freedom ($Nk_BT/V$). The observed strong virial/potential energy correlations do not appear to depend significantly on viscosity. For highly viscous liquids, because of the time scale separation one expects that the slow (relaxing) contributions to pressure are given by the virial and that the slow contributions to energy come from the potential energy. Thus these variables should be highly correlated in their slow equilibrium fluctuations for systems similar to those studied in Ref. \cite{ped07}. This suggests that to a good approximation a single, relaxing parameter controls both quantities, in which case it is obvious to expect that Eq. (\ref{order}) may apply as well.

\begin{figure}\begin{center}
\includegraphics[width=8cm]{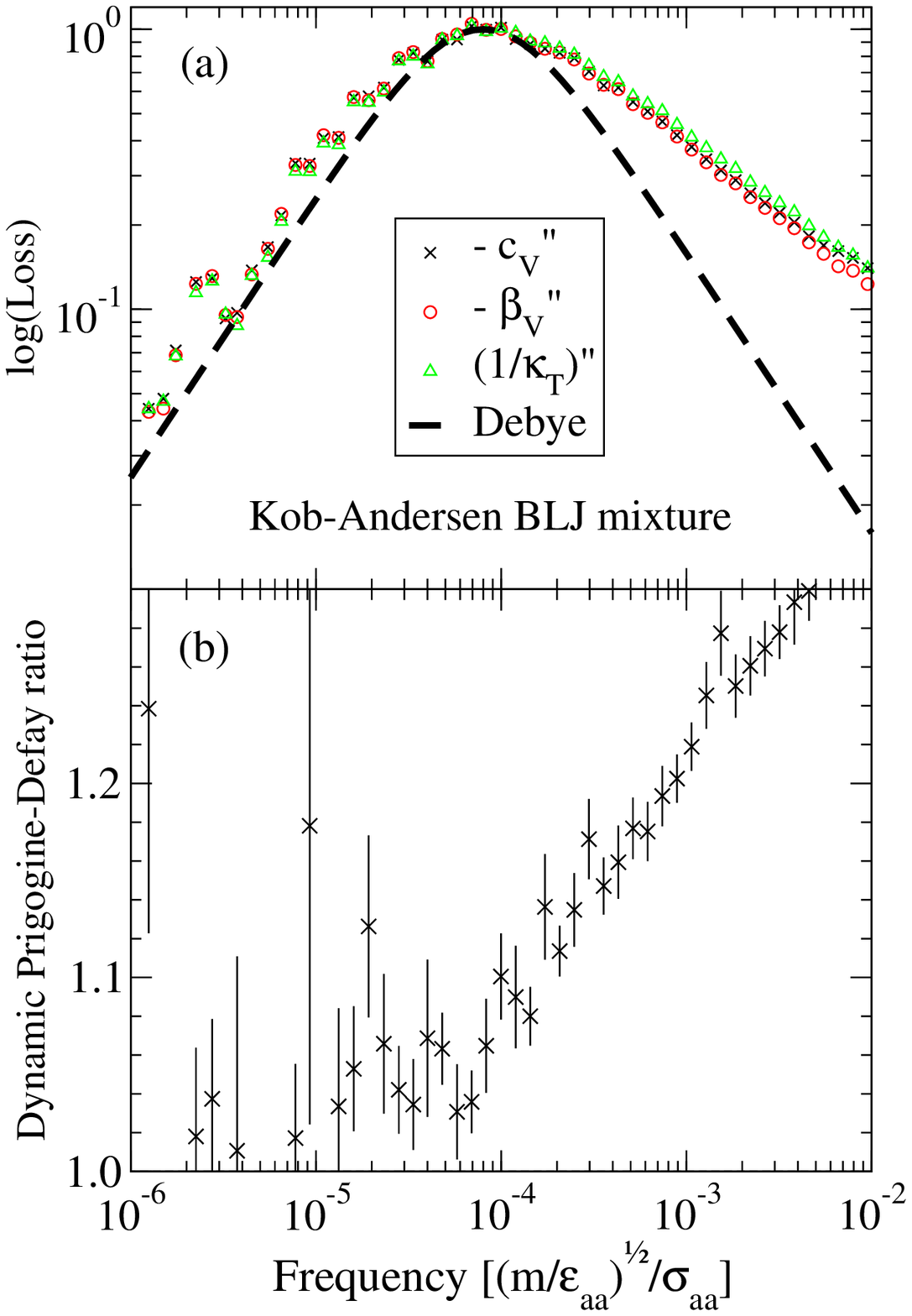}
\includegraphics[width=8cm]{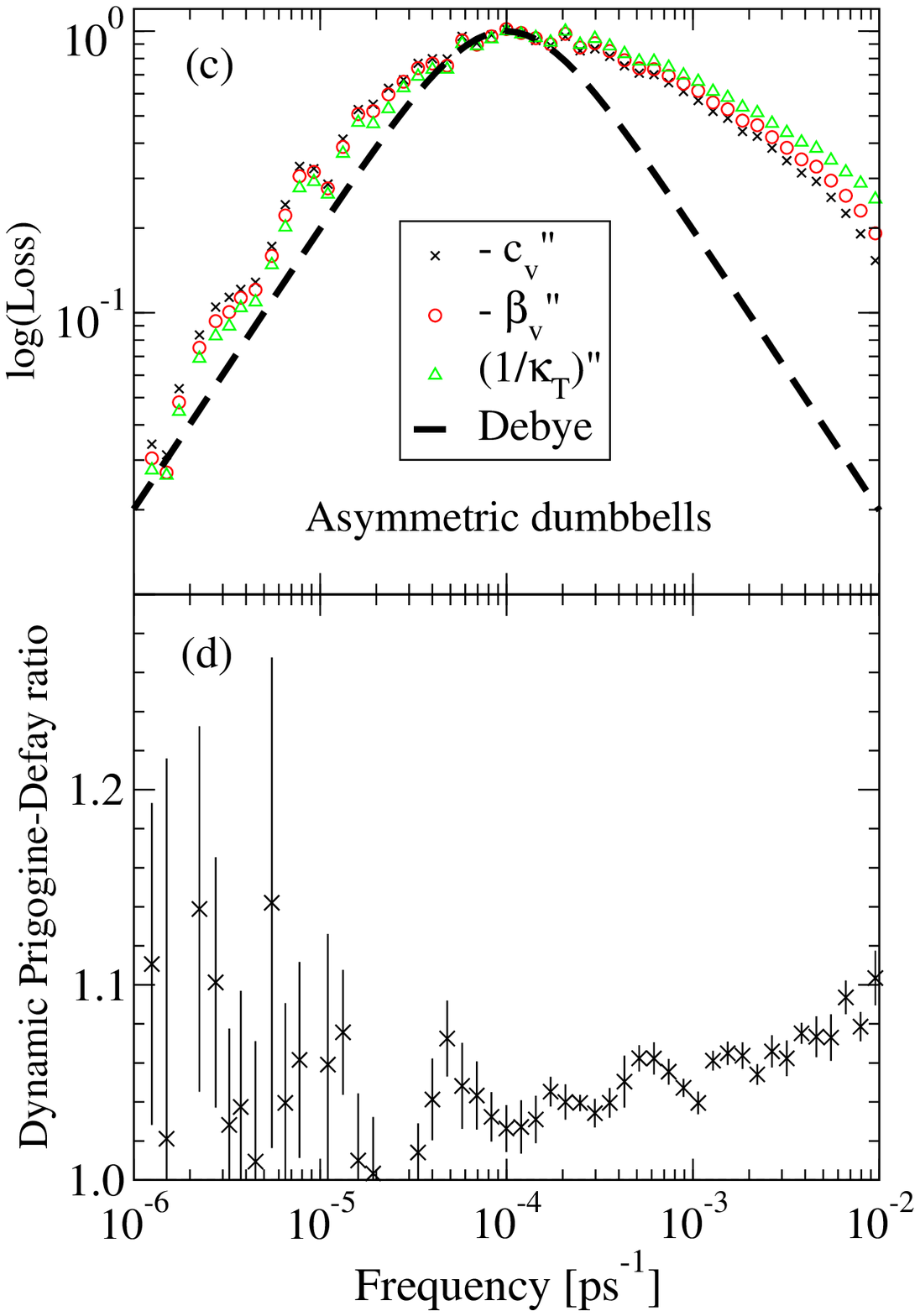}
\caption{(a) show the imaginary parts (scaled to maximum value) of $-c_{v}(\omega)$, $-\beta_v(\omega)$ and $1/\kappa_T(\omega)$ for the Kob-Andersen binary Lennard Jones 80-20 mixture \cite{noteKobAnd}; (c) shows the same for a system of asymmetric dumbbells \cite{note}. The dashed lines indicate 
Debye relaxation with relaxation times $\tau_\alpha=2000$ $\sigma_{aa}\sqrt{m/\epsilon_{aa}}$ and $\tau_\alpha=1600$ ps respectively. 
(b) and (d) show the corresponding dynamic Prigogine-Defay ratios, Eq. (2). For both systems the total simulation time covers more than $10^4\tau_\alpha$. The error bars are estimated from the number of independent simulations.}
\label{figure_pd}
\end{center}\end{figure}

\begin{figure}\begin{center}
\includegraphics[width=8.5cm]{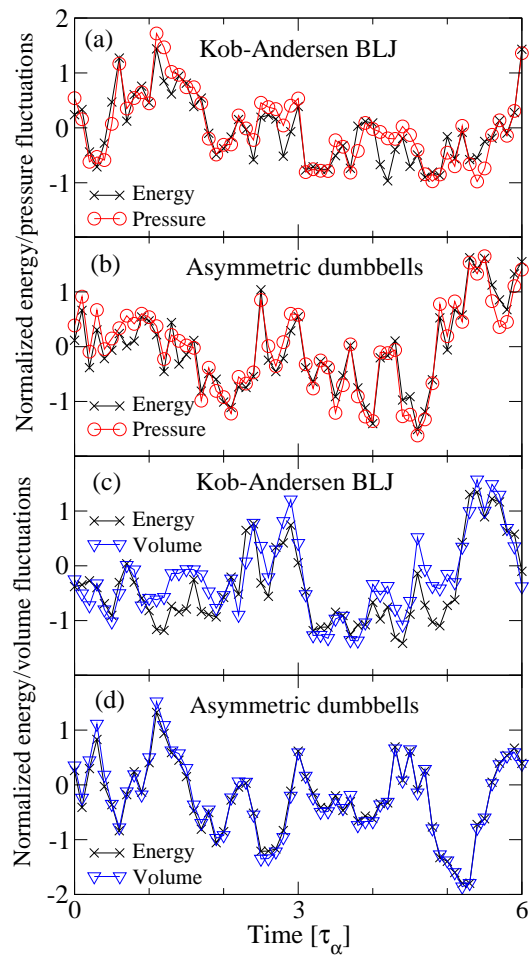}
\caption{(a) and (b): Fluctuations of energy and pressure in the NVT ensemble for the Kob-Andersen binary Lennard-Jones mixture
 \cite{noteKobAnd} and the asymmetric dumbbell system \cite{note}. Each point represents an average over a time interval of $0.1\tau_\alpha$ 
where $\tau_\alpha$ is defined from the loss peak frequency (Figs. 1(a) and 1(c)). Energy and pressure fluctuations are highly correlated, 
showing that a single-order-parameter description is a good approximation. (c) and (d): Fluctuations of energy and volume in the 
NpT-ensemble at the same state points as (a) respectively (b).}
\label{figure_simplepd}
\end{center}\end{figure}

To test this the dynamic Prigogine-Defay ratio was evaluated for two viscous liquid model systems by computer simulation. The test requires that three frequency-dependent thermoviscoelastic response functions must be evaluated. For simulations carried out at constant volume and temperature (the NVT ensemble) the relevant response functions are the isochoric specific heat per unit volume $c_v$, the isothermal compressibility $\kappa_T$, and the ``isochoric pressure coefficient'' $\beta_v\equiv (\partial p/\partial T)_V$. These quantities enter the dynamic Prigogine-Defay ratio \cite{ell07} via

\begin{equation}\label{lambdaTV}
\Lambda_{TV}(\omega)\,=\,-\frac{c''_v(\omega)(1/\kappa_T(\omega))''}{T\left(\beta''_v(\omega)\right)^2}\,.
\end{equation}
The fluctuation-dissipation (FD) theorem \cite{kam81,rei98,nie96,nie99} implies that the above linear response functions may be determined from the equilibrium fluctuations of energy ($E$) and pressure ($p$) as follows. If $\Delta E(t)\equiv E(t)-\langle E \rangle_{V,T}$, $\Delta p(t)\equiv p(t)-\langle p \rangle_{V,T}$, and $\mathcal{C}_\omega\{f(t)\}$ denotes the cosine transform of $f(t)$ at frequency $\omega$, according to the FD theorem

\begin{equation}\label{lambdaTVFD}
\Lambda_{TV}(\omega)\,=\,
\frac{\mathcal{C}_\omega\{\langle \Delta E(0)\Delta E(t)\rangle_{V,T}\}\mathcal{C}_\omega\{\langle \Delta p(0)\Delta p(t)\rangle_{V,T}\}}
{\left(\mathcal{C}_\omega\{\langle \Delta E(0)\Delta p(t)\rangle_{V,T}\}\right)^2}\,.
\end{equation}
If energy and pressure fluctuations correlate perfectly one has $\Delta E(t)\propto \Delta p(t)$ and consequently $\Lambda_{TV}(\omega)=1$ at all frequencies. Equation (\ref{lambdaTVFD}) suggests that the dynamic Prigogine-Defay ratio tests how well the two relevant quantities (here energy and pressure) correlate on the time scale defined by the frequency, it is sort of a time-scale dependent correlation coefficient squared.

Figure 1(a) shows the frequency dependence of the imaginary parts of the three linear-response functions for the standard Kob-Andersen binary Lennard-Jones 80-20 mixture \cite{noteKobAnd}. The three response functions are very similar and all exhibit the characteristic asymmetry towards higher frequencies observed for real glass-forming liquids. Figure 1(b) shows the dynamic Prigogine-Defay ratio; as mentioned this number is unity if and only if the three imaginary parts are strictly proportional \cite{ell07}. This is not the case, but the ratio is fairly close to unity in the frequency range of the main (alpha) process. Thus in the range of frequencies one decade above and below the loss peak frequency -- the ``alpha-relaxation range'' -- the ratio stays below $1.2$. In accordance with the above interpretation of $\Lambda_{TV}(\omega)$ this indicates that for times in the alpha-relaxation range energy and pressure has a correlation coefficient larger than $0.9$.

We also simulated a highly viscous single-component molecular liquid. This is a system of ``asymmetric dumbbell'' molecules defined as two Lennard-Jones spheres of different radii held together by a rigid bond \cite{note}. As shown on Figs. 1(c) and 1(d) the imaginary parts of the response functions are similar, and the dynamic Prigogine-Defay ratio stays below $1.06$ in the alpha-relaxation region.

The results in Fig. 1 show that a class of viscous liquids exists where the dynamic Prigogine-Defay ratio is close to one. This is consistent with earlier computer simulations of the (poorly defined) standard Prigogine-Defay ratio $\Pi$ (Eq. (\ref{1})) for various systems, showing that this quantity is often close to unity \cite{cla79,mor97}. In Figs. 2(a) and (b) we plot energy and pressure equilibrium fluctuations in time for both systems. In order to focus on fluctuations in the alpha time range, both pressure and energy were averaged over one tenth of $\tau_\alpha$ (defined as the inverse loss-peak frequency), corresponding to focussing on the inherent dynamics \cite{sch00} appropriate for understanding the viscous behavior. The correlations are striking. Both systems were also simulated at constant temperature and pressure. Here energy and volume show similarly strong correlations (Figs. 2(c) and (d)). Thus as expected, the appropriateness of a single-parameter description is not ensemble dependent. 

To summarize, it has been shown that it is possible to investigate the single-parameter question by monitoring thermal equilibrium fluctuations. The dynamic Prigogine-Defay ratio provides a convenient test quantity for this, a quantity that is also relevant for experiment because it refers to one single frequency. Clearly, several questions are still unanswered. For instance, it is not known whether there is any link to fragility; one traditionally expects that strong liquids to be single-parameter systems and fragile liquids to require a multi-parameter description. The findings of Ref. \cite{ped07}, however, indicate that network liquids are {\it not} well described by a single parameter in the above sense, whereas purely van der Waals liquids are. Clearly, more work is needed to clarify this and other issues, and we hope that the present results serve to encourage the physics and glass communities to reconsid the old question: ``One or more ``order'' parameters?''

\acknowledgments 
This work was supported by the Danish National Research Foundation's Centre for Viscous Liquid Dynamics ``Glass and Time.''


\begin{thebibliography}{99}

\bibitem{GT_ref} 
W. Kauzmann, Chem. Rev. {\bf 43}, 219 (1948);
G. Harrison, {\it The dynamic properties of supercooled liquids} (Academic Press, New York, 1976);
S. Brawer, {\it Relaxation in Viscous Liquids and Glasses} (American Ceramic Society, Columbus, OH, 1985);
C. A. Angell, K. L. Ngai, G. B. McKenna, P. F. McMillan, and S. W. Martin, J. Appl. Phys. {\bf 88}, 3113 (2000);
C. Alba-Simionesco, C. R. Acad. Sci. Paris (Ser. IV) {\bf 2}, 203 (2001);
P. G. Debenedetti and F. H. Stillinger, Nature {\bf 410}, 259 (2001);
W. Kob, in {\it Slow relaxations and nonequilibrium dynamics in condensed matter, Proceedings of the Les Houches Summer School of Theoretical Physics, Session LXXVII, 1-26 July, 2002}, p. 199, edited by J.-L. Barrat, M. Feigelman, J. Kurchan, and J. Dalibard (Springer, Berlin, 2004); 
J. C. Dyre, Rev. Mod. Phys. {\bf 78}, 953 (2006).%OK

\bibitem{note1} In the glass science the term ``order parameters'' is traditionally used for numbers that in conjunction with pressure and temperature completely characterize a glass, i.e., the molecular structure. This terminology historically preceded the use of the term ``order-parameter'' in the theory for critical phenomena. Nevertheless, in order to avoid confusion it may be prudent to just speak of ``parameters'' or, at least, to write: ``order'' parameters.

\bibitem{pri54} I. Prigogine and R. Defay, {\it Chemical Thermodynamics} (Longman, London, 1954).%OK

\bibitem{dav} R. O. Davies and G. O. Jones, Proc. Roy. Soc. A (London) {\bf 217}, 26 (1952); Adv. Phys. {\bf 2}, 370 (1953).%OK

\bibitem{gol64} M. Goldstein, in {\it Modern aspects of the vitreous state}, Vol. 3, edited by J. D. Mackenzie (Butterworths Scientific, London, 1964), p. 90.%OK

\bibitem{ang76} C. A. Angell and W. Sichina, Ann. N. Y. Acad. Sci. {\bf 279}, 53 (1976).%OK
 
\bibitem{nem95} S. V. Nemilov, {\it Thermodynamic and kinetic aspects of the vitreous state} (CRC, Boca Raton, Florida, 1995).%OK

\bibitem{don01} E. Donth, {\it The glass transition} (Springer, Berlin, 2001).%OK

\bibitem{moy76} C. T. Moynihan {\it et al.}, 1976, Ann. N.Y. Acad. Sci. {\bf 279}, 15 (1976).%OK

\bibitem{kov63} A. J. Kovacs, Fortsch. Hochpolym.-Forsch. {\bf 3}, 394 (1963).%OK

\bibitem{roe77} R.-J. Roe,  J. Appl. Phys {\bf 48}, 4085 (1977).%OK

\bibitem{sch86} G. W. Scherer, {\it Relaxations in Glass and Composites} (Academic, New York, 1986).

\bibitem{moy81} C. T. Moynihan and A. V. Lesikar, Ann. N.Y. Acad. Sci. {\bf 371}, 151 (1981).%OK

\bibitem{nie97} Th. M. Nieuwenhuizen, Phys. Rev. Lett. {\bf 79}, 1317 (1997).%OK

\bibitem{spe99} R. J. Speedy, J. Phys. Chem. B {\bf 103}, 8128 (1999).%OK

\bibitem{ell07} N. L. Ellegaard, T. Christensen, P. V. Christiansen, N. B. Olsen, U. R. Pedersen, T. B. Schr{\o}der, and J. C. Dyre, J. Chem. Phys. {\bf 126}, 074502 (2007).%OK

\bibitem{ric06} R. Richert and S. Weinstein, Phys. Rev. Lett. {\bf 97}, 095703 (2006).%OK - 

\bibitem{nga05} K. L. Ngai, R. Casalini, S. Capaccioli, M. Paluch, C. M. Roland, J. Phys. Chem. B {\bf 109}, 17356 (2005).

\bibitem{dyr03} J. C. Dyre and N. B. Olsen, Phys. Rev. Lett. {\bf 91}, 155703 (2003).

\bibitem{mos04} S. Mossa and F. Sciortino, Phys. Rev. Lett. {\bf 92}, 045504 (2004); see also S. Mossa, E. La Nave, F. Sciortino, and P. Tartaglia, Eur. Phys. J. B {\bf 30}, 351 (2002).%OK

\bibitem{ped07} U. R. Pedersen, N. Bailey, T. B. Schr{\o}der, and J. C. Dyre, cond-mat/0702146 [Phys. Rev. Lett., in press].

\bibitem{kam81} N. G. van Kampen, {\it Stochastical Processes in Physics and Chemistry} (North Holland, Amsterdam, 1981).

\bibitem{rei98} L. E. Reichl, {\it A Modern Course in Statistical Physics}, 2nd Ed. (Wiley, New York, 1998).

\bibitem{nie96} J. K. Nielsen and J. C. Dyre, Phys. Rev. B {\bf 54}, 15754 (1996).%OK

\bibitem{nie99} J. K. Nielsen, Phys. Rev. E {\bf 60}, 471 (1999).%OK

\bibitem{noteKobAnd} The Kob-Andersen binary Lennard-Jones 80-20 mixture was set up [W. Kob and H. C. Andersen, Phys. Rev. Lett. {\bf 73}, 1376 (1994)] with $N=1000$, $V=(9.4\sigma_{aa})^{3}$. The temperature was held constant at $T=0.474$ $\epsilon_{aa}k_{B}^{-1}$ using the Nos\'{e}-Hoover thermostat [S. A. Nos\'{e}, Mol. Phys. {\bf 52}, 255 (1984); W. G. Hoover, Phys. Rev. A {\bf 31}, 1695 (1985)]. The simulations were carried out using Gromacs software [H. J. C. Berendsen, D. van der Spoel, and R. van Drunen, Comp. Phys. Comm. {\bf 91}, 43 (1995); 
E. Lindahl, B. Hess, and D. van der Spoel, J. Mol. Mod. {\bf 7}, 306 (2001)]. The total time covers more than $10^7$ Lennard-Jones time units ($\sigma_{aa}\sqrt{m/\epsilon_{aa}}$).

\bibitem{note} A system consisting of 512 asymmetric dumbbell molecules modelled as two Lennard-Jones spheres connected by a rigid bond was simulated. The dumbbells were parameterized to mimic toluene. A large sphere (mimicking a phenyl group) was taken from the Wahnstr\"om OTP model [L. J. Lewis and G. Wahnstr\"om, Phys Rev. E {\bf50}, 3865 (1994)] with the parameters $m_p=77.106$ u, $\sigma_p=0.4963$ nm and $\epsilon_p=5.726$ kJ/mol. A small sphere (mimicking a methyl group) was taken from UA-OPLS [W. L. Jorgensen, J. D. Madura and Carol J. Swenson, J. Am. Chem. Soc. {\bf106}, 6638 (1984)] having $m_m=15.035$ u, $\sigma_m=0.3910$ nm and $\epsilon_m=0.66944$ kJ/mol. The bonds were kept rigid using the LINCS  algorithm [B. Hess, H. Bekker, H. J. C. Berendsen and J. G. E. M. Fraaije, J. Comp. Chem. {\bf18}, 1463 (1997)] with a bond length of $d=0.29$ nm. The volume was $V=77.27$ nm$^3$ giving an average pressure of approximately 1 atm. The temperature was held constant at $T=130$ K using the Nos\'{e}-Hoover thermostat. The simulations were carried out using Gromacs software.

\bibitem{cla79} C. A. Angell, J. H. R. Clarke, and L. V. Woodcock, Adv. Chem. Phys. {\bf 48}, 397 (1981).%OK

\bibitem{mor97} D. Morineau, G. Dosseh, R. J. M. Pellenq, M. C. Bellissent-Funel, and C. Alba-Simionesco, Molecular Simulations {\bf 20}, 95 (1997).

\bibitem{sch00} T. B. Schr{\o}der, S. Sastry, J. C. Dyre, and S. C. Glotzer, J. Chem. Phys. {\bf 112}, 9834 (2000).

\end{thebibliography}
\end{document}